\documentclass[aps,twocolumn,showpacs,pra,superscriptaddress]{revtex4-2}
\usepackage{graphics}
\graphicspath{ {images/} }
\usepackage{mathtools}
\usepackage{bm}
\usepackage{dsfont,amsthm,amsbsy}
\usepackage{verbatim}
\usepackage{amssymb}
\usepackage{amsmath}
\usepackage{comment}
\usepackage{bbm}
\usepackage{overpic}
\usepackage{graphicx}
\usepackage{epstopdf}
\usepackage{subfigure}
\usepackage{natbib}
\usepackage{epsfig}
\usepackage{amsfonts}
\usepackage{mathrsfs}
\usepackage{sidecap}
\usepackage{lipsum}
\usepackage{url}
\usepackage[toc,page,title,titletoc,header]{appendix}
\usepackage[colorlinks,linkcolor=blue,citecolor=blue,anchorcolor=blue,urlcolor=blue]{hyperref}
\usepackage{hyperref}
\usepackage{resizegather}
\usepackage{tikz}
\usetikzlibrary{arrows.meta}
\usepackage{float}
\usepackage{mathbbol}
\usepackage[normalem]{ulem}
\usepackage{cancel}
\usepackage{upgreek}
\usepackage{enumitem}

\newcommand{\bl}{\begin{aligned}}
\newcommand{\el}{\end{aligned}}

\def\be{\begin{equation}}
\def\ee{\end{equation}}

\def\bi{\begin{itemize}}
\def\ei{\end{itemize}}
\def\bn{\begin{enumerate}}
\def\en{\end{enumerate}}
\def\bea{\begin{eqnarray}}
\def\eea{\end{eqnarray}}
\def\no{\nonumber}
\def\ba{\begin{array}}
\def\ea{\end{array}}
\def\bd{\begin{displaymath}}
\def\ed{\end{displaymath}}

\def\ket#1{\left|#1\right\rangle}

\begin{document}
	
	\title{Dissipation-Induced Deviations from Kibble-Zurek Scaling in\\
		Non-Hermitian Quantum Annealing}
	
	
	\author{H. Najafzadeh}
	\email{honeynjf@gmail.com}
	\thanks{Equal contribution}
	\affiliation{Department of Physics, Sharif University of Technology, Tehran 1458889694, Iran}
	
	\author{S. Sadeghizade}
	\email{sajadsadeghizadeh@gmail.com}
	\thanks{Equal contribution}
	\affiliation{Department of Physics, Sharif University of Technology, Tehran 1458889694, Iran}
	
	\author{R. Jafari}
	\email{raadmehr.jafari@gmail.com}
    \affiliation{Department of Physics, Institute for Advanced Studies in Basic Sciences (IASBS), Zanjan, Iran}
	\affiliation{School of Quantum Physics and Matter, Institute for Research in Fundamental Sciences (IPM), Tehran, Iran}

	\author{A. Langari}
	\email{abdollah.langari@gmail.com}
	\affiliation{Department of Physics, Sharif University of Technology, Tehran 1458889694, Iran}

\begin{abstract}
We revisit the quantum annealing problem in the non-Hermitian transverse-field Ising model. We determine, both analytically and numerically, the intrinsic transition probabilities and the resulting defect density. Our results reveal that, unlike the Hermitian case where defect production is dominated by modes near the gap-closing point, the non-Hermitian dynamics involve significant contributions from broad momentum sectors. We find that, depending on the dissipation strength, the defect density exhibits standard Kibble-Zurek scaling, anti-Kibble-Zurek behavior, and a suppression faster than the Kibble-Zurek prediction. We demonstrate that these deviations from the standard Kibble-Zurek scaling can be understood in terms of the underlying excitation probabilities. Specifically, the fast decay of the defect density originates from a vanishing excitation probability spanning a range of annealing times across all allowed modes, even at the gap-closing points. In contrast, the anti-Kibble-Zurek behavior arises from supplementary excitations facilitated by dissipation over a broad range of allowed modes, particularly those situated away from the gap-closing region.
\end{abstract}

\maketitle

\section{Introduction}

Formulating a comprehensive theoretical framework for nonequilibrium phenomena constitutes a fundamental challenge in modern physics~\cite{Hohenberg1977,Chou2011}. Central to this endeavor is a rigorous understanding of adiabatic dynamics and the mechanisms underlying its breakdown. Such insights are of paramount importance for the advancement of quantum technologies, which demand the precise coherent manipulation of complex many-body systems.
Quantum annealing (QA)~\cite{Grabarits2025,Subires2022,Bando2020,Liu2015,Suzuki2010} and adiabatic quantum computation provide a broadly studied paradigm for preparing low-energy states of interacting many-body Hamiltonians and, more generally, for encoding optimization tasks into ground-state search problems~\cite{Farhi2000,Albash2018}. 
However, suppressing excitations becomes challenging when a control parameter of the system varies across the quantum critical point (QCP) where the energy gap closes~\cite{Dziarmaga2005,Polkovnikov2011,Grabarits2025b,Sen2008,Kolodrubetz2012,Shreyoshi2009}. As the system crosses the critical point, excitations are inevitably created, challenging the possibility of achieving perfectly adiabatic driving.
Within this context, the Kibble-Zurek mechanism (KZM), which elucidates the dynamics across a continuous phase transition, provides a relatively simple theoretical framework and valuable insights into connecting nonequilibrium critical dynamics with equilibrium criticality~\cite{Dziarmaga2005,Polkovnikov2011,Grabarits2025b,Sen2008,Kolodrubetz2012,Shreyoshi2009,Kibble1976,Zurek1985,Grandi2011,Kou2022,Kou2023,Jara2026}.

The central prediction of the KZM is that the average topological defect density, $n$, displays a universal power-law scaling as a function of the annealing time $\tau$ i.e., $n=\tau^{-\beta}$. This universal behavior is determined by the equilibrium 
critical exponents of the phase transition i.e., $\beta=d\nu/(1+z\nu)$, where, $\nu$ and $z$ are the correlation length and 
dynamic critical exponents, respectively and $d$ is dimensionality of the system~\cite{Dziarmaga2005,Polkovnikov2011,Dario2008,Dario2009}. 
However, a number of studies on ferroelectric phase transitions \cite{Griffin2012}, polar phase of $^3$He \cite{Rysti2021} and on open quantum systems \cite{Nigmatullin2016,Anirban2016,Puebla2020,Sadeghizade2025,Singh2021,Singh2023,Gao2017,Jafari2026a,Balducci2023,Bermudez2009,Jafari2026b,Kou2026,Kou2025} have reported deviations from the conventional Kibble-Zurek scaling. The observations correspond to an increase in the excitation density due to an enhanced annealing time, known as anti-Kibble-Zurek (AKZ) behavior.

Furthermore, in realistic devices and engineered platforms, dissipation, gain/loss and measurement backaction can significantly impact state preparation. This motivates annealing protocols in open-system settings in terms of non-Hermitian generators \cite{Breuer2002,Rivas2012,Ashida2020,Bergholtz2021,Qiu2019}. 
A non-Hermitian system can exhibit exceptional points, where eigenvalues and eigenvectors coalesce, resulting in a loss of completeness and leading to qualitative modifications of adiabatic response and nonadiabatic production. This phenomenon can be extended to non-Hermitian Kibble-Zurek physics \cite{Dora2019}.
These modifications are characterized by scaling and crossover structures that are linked to complex gaps, exceptional-point geometry and quantum-metric-induced defect freezing~\cite{Ashida2020,Bergholtz2021,Jing2024,Zhou2018,Zhou2021,Nehra2024,Sim2023}.
%

%

In this work, we revisit the non-Hermitian quantum annealing problem in the one-dimensional ferromagnetic Ising chain and provide analytical and numerical evaluations of intrinsic transition probabilities and the resulting defect density. 
It has been reported that such a protocol in the ferromagnetic Ising chain decreases defect density and can enhance the probability of reaching the ground state while reducing the effective annealing time~\cite{Nesterov2013,Brody2014,Oshiyama2020}. 
We have demonstrated that, in contrast to results of previous work~\cite{Nesterov2013}, the modes away from the gap-closing mode have a significant contribution to the defect density. In other words, the approximation that effectively restricts the dynamics to modes close to the gap-closing mode is not generically justified in non-Hermitian annealing.
It has been demonstrated that, in the presence of dissipation, the excitation probability is non-zero around the gap-closing mode for short annealing times. However, we show that there exists a range of annealing time intervals over which the excitation probability is minimal for all permitted modes. Beyond this interval, the excitation probability becomes significant for a broad range of permitted modes, excluding the gap-closing mode. Consequently, depending on the dissipation strength, the defect density exhibits Kibble-Zurek scaling and anti-Kibble-Zurek behavior, as well as suppression faster than the Kibble-Zurek prediction. 

Our findings provide new insights into an important paradigm in nonequilibrium statistical physics. Moreover, since adiabatic driving of quantum many-body systems is relevant to adiabatic quantum computation, these results may offer new perspectives on approaches that circumvent the KZM scaling law and enable adiabatic dynamics~\cite{Doria2011,Adolfo2012,Campbell2015,Puebla2020b,Mathey2010}.

In this article, the underlying model is introduced in Sec.~\ref{sec:model} and the analytic solution of the quench dynamics is presented in Sec.~\ref{sec:mode_dynamics}. Several paths of a linear quench protocol are then presented in Sec.~\ref{sec:protocols} to investigate defect production during the annealing process. Finally, in Sec.~\ref{sec:conclusions}, the findings are summarized and concluded.

\section{Model and exact solution}\label{sec:model}

The Hamiltonian of an $N$-site one-dimensional transverse-field Ising chain with dissipation is given as~\cite{Nesterov2013}

%
\begin{equation}
	H = -\frac{J}{2}\sum_{n=1}^{N}
	\left[
	g\,\sigma_n^{x}
	+
	\sigma_n^{z}\sigma_{n+1}^{z}
	+
	i\,2\delta\,\sigma_n^{-}\sigma_n^{+}
	\right],
	\label{eq:H_spin}
\end{equation}
%
where $\sigma_n^{\alpha}$ are Pauli matrices and $\sigma_n^{\pm}=(\sigma_n^{z}\pm i\sigma_n^{y})/2$ are the spin raising and lowering operators.  
Periodic boundary conditions, $\sigma_{N+1}^{\alpha}\equiv\sigma_1^{\alpha}$, are imposed.  
Here, $J>0$ denotes the nearest-neighbor ferromagnetic coupling and $g$ is the transverse field.  
The dissipative term proportional to $\delta$ accounts for the gain ($\delta<0$) or loss ($\delta>0$) of the spin-down state $|\downarrow\rangle$ during interaction with the environment at a rate $\delta$~\cite{Naji2022}.  
The term ``dissipative'' refers to the tunneling of intrinsic system states into a continuum, leading to an effective non-Hermitian description that naturally arises within the Feshbach projection formalism~\cite{Feshbach1958,Feshbach1962}.

The Hamiltonian, Eq.~(\ref{eq:H_spin}), can be mapped to the free spinless fermion model with complex chemical potential~\cite{Nesterov2013} by Jordan-Wigner transformation~\cite{Lieb1961,Pfeuty1970,Barouch1971} (see Appendix~\ref{app:JW})

%
\bea
\bl
		H
		= -\frac{J}{2}\sum_{n=1}^{N}
		\Big[
		& c_n^\dagger c_{n+1}
		+ c_{n+1}^\dagger c_n
		+ c_n^\dagger c_{n+1}^\dagger
		+ c_{n+1} c_n\\
\label{eq:H_pm_main}
		& - 2\tilde{g}\, c_n^\dagger c_n
		+ \tilde{g} + i\delta
		\Big],
\el
\eea
%
%
\begin{figure}[t]
	\centering
	\includegraphics[width=0.38\textwidth]{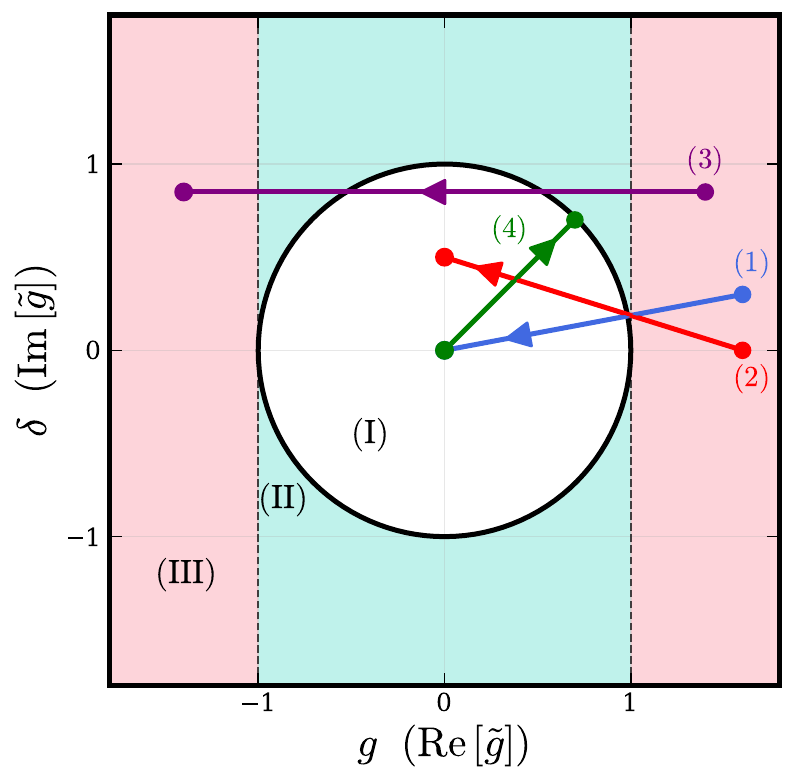}
	\caption{(Color online) Phase diagram of the dissipative transverse field Ising model in the complex control-parameter ($\tilde {g}=g+i\delta$) plane.
		The thick black circle denotes the exceptional ring $g^2+\delta^2=1$ (zero energy gap). 
		Region (I): inside the exceptional ring $g^2+\delta^2<1$ (pure real spectrum gap).
		Region (II): outside the exceptional ring with $|g|<1$ (pure imaginary spectrum gap). 
		Region (III): outside the exceptional ring with $|g| >1$ (complex gap).
         Four solid lines, are displayed in different colors, represent four different quench paths. The quench directions
         are indicated by the arrows (annealing trajectories).
	}
	\label{fig:phase_space}
\end{figure}
%
where $\tilde g=g+ i\delta$ and $c_n^\dagger$ ($c_n$) is the spinless fermion creation (annihilation) operator.
Performing a Fourier transformation, $c_n = (e^{-i\pi/4}/\sqrt{N})\sum_k c_k e^{i n k}$ (with the phase
factor $e^{-i\pi/4}$ added for convenience) and imposing antiperiodic fermionic boundary conditions (even parity sector), $c_{N+1}=-c_1$
(which leading to $k = 2\pi p/N$, with $p=1/2,3/2,\ldots, N-1/2$),
$H(t)$ is expressed as a sum over decoupled mode Hamiltonians: 
$H(t) = \sum_{k>0} H_k(t)$,
\begin{equation}
H_k(t)=\Psi_k^\dagger \mathcal{H}_k(t) \Psi_k - \varepsilon_0(k) \mathbb{1},
\end{equation}
with $\varepsilon_0(k)=J (\cos(k) + i \delta)$ and, 
%
\begin{equation}
	\mathcal{H}_k(t)
	= J
	\begin{pmatrix}
		\tilde g - \cos(k) & \sin(k) \\
		\sin(k)           & -\tilde g + \cos(k)
	\end{pmatrix},
	\label{eq:BdG_matrix}
\end{equation}
%
where $\Psi_k=(c_k \; c_{-k}^\dagger)^{T}$ are Nambu spinors.
The corresponding complex quasiparticle energy is given as~\cite{Nesterov2013} (see Appendix~\ref{app:Bogoliubov})
%
\begin{equation}
	\varepsilon_k
	= \pm J\sqrt{(\tilde g-\cos(k))^2+\sin^2k},
	\label{eq:Energy_k}
\end{equation}
%
which is gapless if the imaginary and real part of $\varepsilon_k$ become zero ($Im(\varepsilon_k)=0, Re(\varepsilon_k)=0$). This leads to
%


\bea
\label{eq6}
&&k^{\ast}=\arccos(g_c),\\
\label{eq7}
&&g_c^2+\delta^2=1.
\eea
%
where $g_c$ is the critical value of $g$. Eq.~(\ref{eq6}) implies a limitation $|g_c| \leq 1$, while Eq.~(\ref{eq7}) represents the exceptional ring (degeneracy points of the non-Hermitian Hamiltonian). 
Consequently, the system can be divided into three regions as illustrated in Figure~\ref{fig:phase_space}.
%

In the following sections, we will study the dynamics of the system for different quench paths across the exceptional ring along a linear protocol, as represented in the phase diagram in Figure~\ref{fig:phase_space}.

%
\section{Time-dependent Hamiltonian}
\label{sec:mode_dynamics}

To study the dynamics of non-Hermitian transverse field Ising model, we consider that the transverse magnetic field and dissipation rate are time-dependent. 
In such a case, the time-dependent Hamiltonian $H(t)$ in Eq. (\ref{eq:H_spin}) can still be expressed as a sum of decoupled time-dependent Hamiltonian $\mathcal{H}_k(t)$, given in Eq. (\ref{eq:BdG_matrix}) with time-dependent transverse field $g(t)$ and dissipation $\delta(t)$.
The time evolution of a generic state is governed by the time-dependent Schr\"odinger equation
\begin{equation}
	i\frac{d}{dt}\ket{\psi_k(t)} = H_k(t)\ket{\psi_k(t)}.
	\label{eq:Sch_generic}
\end{equation}
In the even-parity subspace spanned by
$\{\ket{0},\,c_k^\dagger c_{-k}^\dagger\ket{0}\}$, where $\ket{0}$ is the vacuum, we parametrize the
state as
\begin{equation}
	\ket{\psi_k(t)} =
	\Big[
	C_{1,k}(t)\ket{0}
	+ C_{2,k}(t)\,c_k^\dagger c_{-k}^\dagger\ket{0}
	\Big]
	\exp\!\left[i\int_{0}^{t}\varepsilon_0\big(k,t'\big)\,dt'\right],
	\label{eq:psi_k_with_phase}
\end{equation}

Inserting Eq.~\eqref{eq:psi_k_with_phase} into the Schr\"odinger equation
\eqref{eq:Sch_generic} and using Eq.~\eqref{eq:BdG_matrix}, the
amplitudes satisfy a two-level problem governed by the traceless
Bogoliubov-de Gennes matrix \cite{Lieb1961,Pfeuty1970,Nesterov2013}:
\begin{equation}
	i\frac{d}{dt}
	\begin{pmatrix}
		C_{1,k}(t)\\[1mm] C_{2,k}(t)
	\end{pmatrix}
	=J
	\begin{pmatrix}
		\tilde g(t)-\cos(k) & \sin(k)\\[1mm]
		\sin(k)    & -\tilde g(t)+\cos(k)
	\end{pmatrix}
	\begin{pmatrix}
		C_{1,k}(t)\\[1mm] C_{2,k}(t)
	\end{pmatrix}.
	\label{eq:Sch_2level}
\end{equation}
In the following, we will omit the momentum label on the
amplitudes and simply write $C_{1}(t)$ and $C_{2}(t)$.

Equation~\eqref{eq:Sch_2level} with the linear time-dependent transverse field and linear time-dependent dissipation is exactly solvable (see Appendix~\ref{app:weber}).
It is straightforward to show that the state of the system at time $t$ can be expressed as a linear combination of the right-eigenvectors ($|u_{\pm,k}(t)\rangle$) of the instantaneous eigenstates of the Hamiltonian (see Appendix \ref{app:Bogoliubov})
%
\begin{equation}
	\ket{\psi_k(t)} =\alpha_k |u_{-,k}(t)\rangle + \beta_k |u_{+,k}(t)\rangle,
	\label{adibatic_expansion}
\end{equation}	
%
where
%
\begin{equation}
	\begin{align}
	\alpha_k &= e^{i\int_{0}^{t}\varepsilon_0(k,t')\,dt'} \Big( C_2(T)\cos\!\frac{\theta_k}{2} - C_1(T)\sin\!\frac{\theta_k}{2}\Big),\\
\no
	\beta_k  &= e^{i\int_{0}^{t}\varepsilon_0(k,t')\,dt'} \Big( C_2(T)\sin\!\frac{\theta_k}{2} + C_1(T)\cos\!\frac{\theta_k}{2}\Big).
	\label{eq:alpha_beta_map_main}
	\end{align}
\end{equation}
%
Here, $\tan(\theta_k)=\sin(k)/(\tilde g(t)-\cos(k))$, $T=\sqrt{a_k}\,(t-t_{0,k})$, $a_k =J\dot{\tilde g}(t_{0,k})$ and $t_{0,k}$ is complex avoided-crossing time, i.e.,
$\tilde g(t_{0,k})-\cos(k)=0$  (see Appendix~\ref{app:Bogoliubov}).\\

The excitation probability that remains well-defined even for non-unitary evolution (complex \(\tilde g\)) is expressed as given in Eq.~\eqref{eq:Pk_num_correct}, which reduces to the following form whenever the Hamiltonian is Hermitian at the end of the quench [as in path~($1$) in Fig.\ref{fig:phase_space}],
%
\begin{equation}
	P_k(t)=\frac{|\beta_k|^2}{|\alpha_k|^2+|\beta_k|^2}.
	\label{eq:Pk_normalized}
\end{equation}
%
This leads to the defect (excitation) density of the system at time $t$
%
\begin{equation}
	n(t)=\frac{2}{N}\sum_{k>0} P_k(t).
	\label{eq:n_def}
\end{equation}
%
Equivalently, the Schr\"odinger  Eq.~(\ref{eq:Sch_2level}) can be solved numerically to obtain the defect density.
The approach of numerical simulation has been elaborated in Appendix~\ref{subsec:numerics}, which requires the subtle 
computation of right and left eigenvectors and the determination of the probability of excitations.


%
\begin{figure}[!t]
\centerline{
\includegraphics[width=0.49\linewidth]{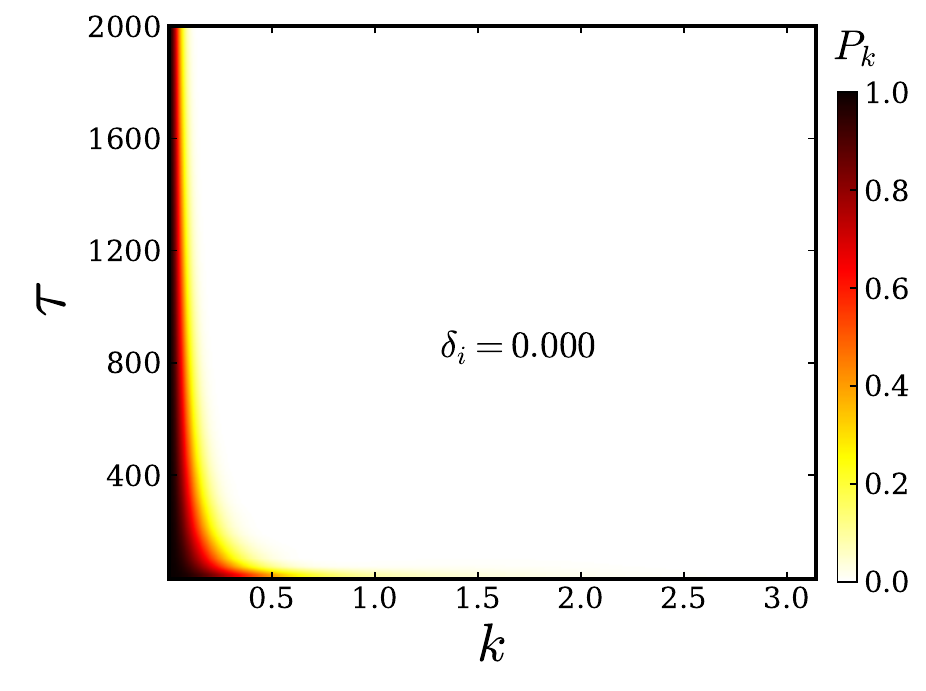}
\includegraphics[width=0.49\linewidth]{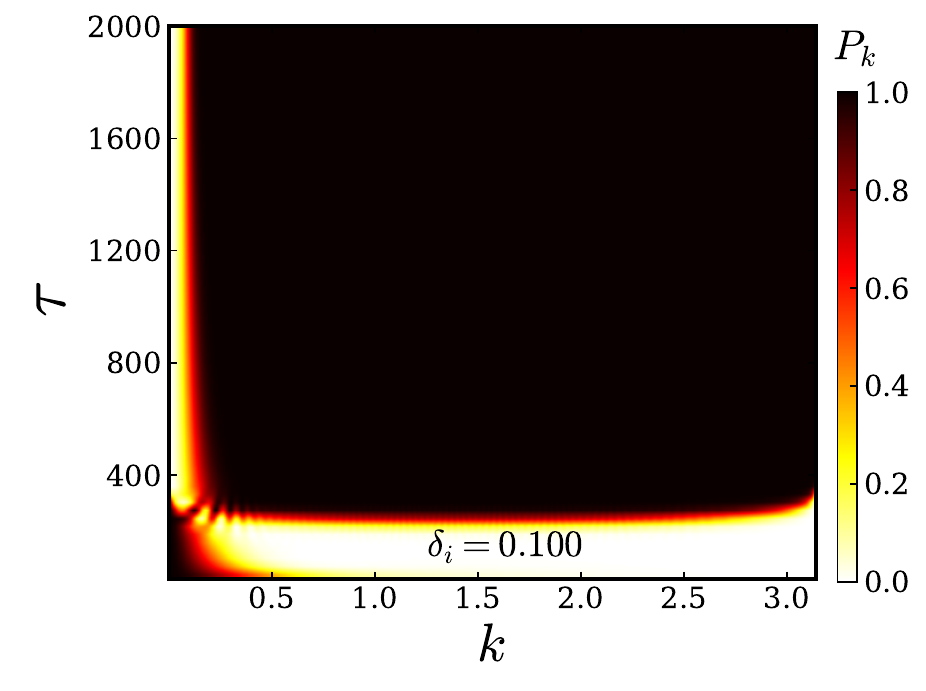}}
\caption{(Color online) The density plot of excitation probability at the end 
of quench $P_k(\tau)$ vs $k$ and $\tau$ for path~($1$) with Hamiltonian parameters $J=0.5$, $g_i=10$, and $N=1024$. The left panel represents $P_k(\tau)$ 
for the unitary evolution of the model ($\delta=0$). 
In the right panel $P_k(\tau)$ has been depicted for the non-Hermitian evolution of 
the system for dissipation strength $\delta=0.1$ for the quench path ($1$).}
\label{fig:P1_heatmap}
\end{figure}
%

\section{LINEAR QUENCH PATHS}\label{sec:protocols}
In this section, we study the dynamics of the system under various linear quench paths after quenching it across the exceptional point. 
All quench paths are indicated in the phase diagram in the ($g$, $\delta$) plane (Figure~\ref{fig:phase_space}).
We demonstrate that the presence of dissipation leads to strikingly different behaviors of defect density following fast and slow quenches.


\subsection{The quench path ($1$): radial linear ramp}\label{subsec:protocol1}

As a preliminary, we consider the radial linear ramp in the complex plane,
following the quenching protocol~\cite{Nesterov2013},
%
\begin{equation}
	\tilde{g}(t) = \lambda(\tau - t),
	\qquad 0 \le t \le \tau,
	\label{protocol_1}
\end{equation}
%
with a constant complex rate $\lambda=\tilde g(0)/\tau$, from an initial value $\tilde g(0)=g_i+i\delta_i$ at $t_i=0$, 
to a final value $\tilde g(\tau)=0$ at $t_f=\tau$.
This trajectory is represented by the blue arrow [quench path ($1$)] in Fig.~\ref{fig:phase_space} and provides a
direct non-Hermitian analog of a linear annealing schedule~\cite{Nesterov2013,Ashida2020,Bergholtz2021}.
In such a case, the Hamiltonian $\mathcal{H}_k(t)$ in Eq. (\ref{eq:BdG_matrix}) can be mapped to the time-dependent Schr\"{o}dinger equation 
of a generalized Landau-Zener problem with complex crossing time \(\Delta_k(t_{0,k})=0\), which is exactly solvable~\cite{Zener1932,Vitanov1999,VitanovGarraway1996}.

%
\begin{figure}[t]
	\centering
	\includegraphics[width=0.98\linewidth]{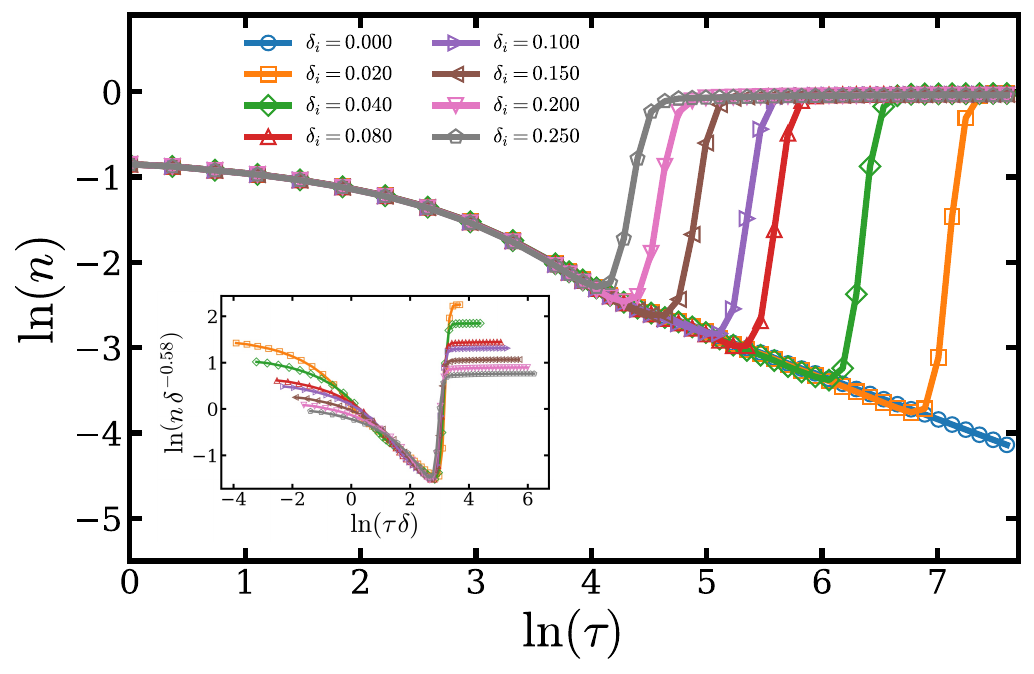}
	\caption{(Color online) The defect density vs annealing time 
     for the quench path ($1$), represented in phase diagram Fig.~\ref{fig:phase_space},
	with Hamiltonian parameters $J=0.5$, $g_i=10$, and $N=1024$ for several values of dissipation ($\delta_i$).
	Inset: 	The scaling behavior of $n(\tau)$ near the minimum is observed by collapsing all curves of various $\delta$ by plotting $n(\tau)/\delta^{\gamma_1}$ vs $\tau/\delta^{-\alpha_1}$.}
	\label{fig:P1_n_tau}
\end{figure}
%

For non-unitary evolution, the excitation probability at time $t$ ($P_k(t)$) is quantified by a normalized projection~\cite{Nesterov2013,Ashida2020} given in Eq.~(\ref{eq:Pk_normalized}) [see also Eq.~\eqref{eq:Pk_num_correct}].
The density plot of excitation probability at the end of quench $P_k(\tau)$, has been plotted versus $k$ and the annealing time $\tau$ in Fig.~\ref{fig:P1_heatmap}. 
The left panel displays $P_k(\tau)$ of the well-known unitary evolution of the system ($\delta=0$), where the modes close to the gap-closing mode ($k \sim 0$) are only excited. 
In the right panel $P_k(\tau)$ represented for the non-Hermitian evolution of the model for dissipation strength $\delta=0.1$.
As clearly evident, in the presence of dissipation, the excitation probability of modes away from the gap-closing mode becomes non-zero as the annealing time $\tau$ increases. In other words, in contrast to unitary evolution ($\delta=0$), the system undergoes nonadiabatic evolution even for modes away from the gap-closing mode and can affect the behavior of defect density. 

%
\begin{figure}[!t]
\centerline{
\includegraphics[width=0.95\linewidth]{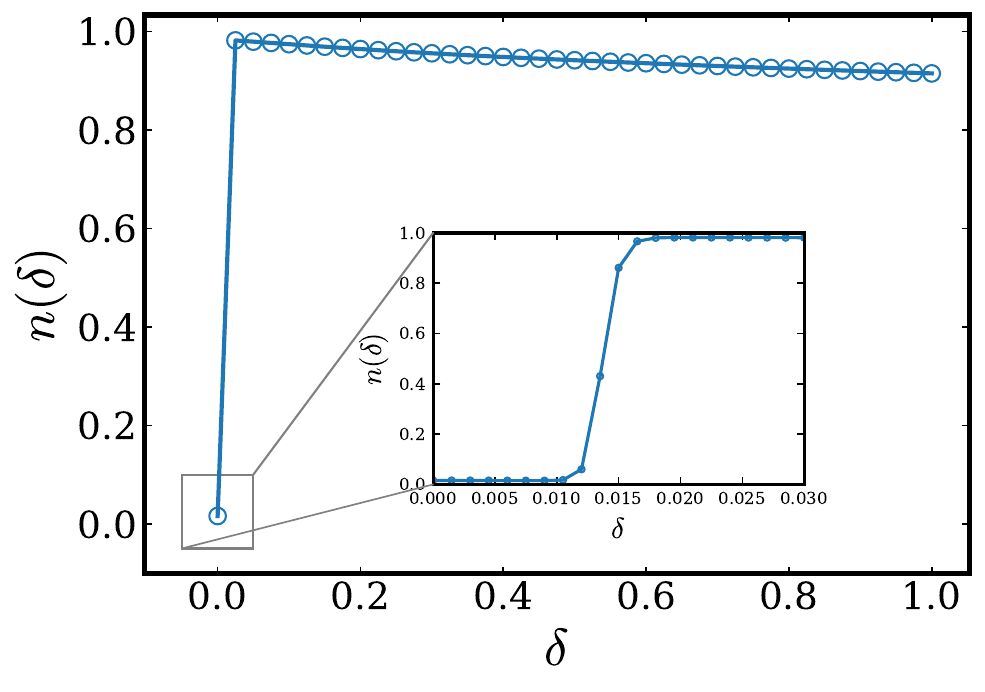}}
\caption{(Color online)  The density of defects vs dissipation parameter ($\delta$) for quench path~(1).
(Inset) Zoomed plot for very small values of $\delta$, which shows the sudden change in density upon increasing dissipation.}
\label{fig:n_vs_delta_P1}
\end{figure}


The density of defects $n(\tau)$ for the quench path given in Eq. (\ref{protocol_1}), has been plotted versus annealing time $\tau$ in Fig.~\ref{fig:P1_n_tau} for 
several values of dissipation strength. It can be clearly seen that, for short annealing time, the effect of dissipation is negligible and 
the density of defects scales as a power law in agreement with the KZM prediction $n(\tau)\propto\tau^{-1/2}$ ($\delta=0$)~\cite{Dziarmaga2005}.
However, for longer annealing time, dissipation-induced effects dominate the nonadiabatic dynamics, leading to the growth of $n(\tau)$ 
with the annealing time $\tau$. This is indicative of a crossover to an AKZ regime, where increasing the annealing time results in a higher excitation of the system. 
In the very long annealing times, $n(\tau)$ is completely governed by the AKZ contribution and makes a plateau versus annealing time.

Thus the defect generation appears to be controlled by two competing processes: (i) non-adiabatic excitations, suppressed by increasing the annealing time
as in a KZM, and (ii) dissipation-induced excitations, amplified by increasing $\tau$ as in an AKZ scenario.
The competition between the two processes yield local minima of $n(\tau)$ at time scales $\tau^{(1)}_{opt}$, acting as
optimal times and as the dissipation strength increases, the optimal annealing time decreases.

A more detailed analysis reveals that the optimal annealing time scales in a power law manner with dissipation, $\tau^{(1)}_{opt} \sim \delta^{-\alpha_1}$, where $\alpha_1 = 1 \pm 0.05$.
Furthermore, a similar linear scaling is observed for the minimum defect density, which occurs at $\tau^{(1)}_{opt}$, i.e., $n^{(1)}_{min} \sim \delta^{\gamma_1}$, with $\gamma_1 = 0.58 \pm 0.05$.
These findings suggest a scaling behavior of $n(\tau)$ close to the minimum, i.e., $n(\tau)$ is invariant under the scaling transformations $n(\tau) \rightarrow n(\tau)/n^{(1)}_{min}$ and $\tau \rightarrow \tau/\tau^{(1)}_{opt}$.
The scaling of defect density corresponding to various values of $\delta$ is illustrated in the inset of Figure~\ref{fig:P1_n_tau}, which demonstrates that all curves collapse into a single graph under scaling. These scaling functions represent the promised universality of $n(\tau)$ in the face of dissipation.

It should be noted that, while the modes away from the gap-closing ones have a significant contribution to the defect density, this contribution has been ignored in previous work~\cite{Nesterov2013}. Considering the significant contribution of all modes (as illustrated in the right panel of Figure~\ref{fig:P1_heatmap}), the defect density exhibits a sudden increase upon adding dissipation to the system.
In Figure~\ref{fig:n_vs_delta_P1}, the defect density is plotted versus $\delta$ for $\tau=2000$. As observed, the defect density increases with increasing dissipation, which is in contrast to the results presented in Ref.~\cite{Nesterov2013}.

%
In the following we introduce alternative quench paths designed
to realize a controlled suppression of excitations by dissipation.


\subsection{The quench path ($2$): increasing dissipation enhancing adiabaciticity}\label{subsec:protocol2}

In the quench path ($2$), the complex field is linearly changed from an initial pure real value $\tilde g(0) = g_i$ at $t_i = 0$ to a pure imaginary value $\tilde g(\tau) = i \delta_f$ at $t_f = \tau$ (red solid arrow in Fig.~\ref{fig:phase_space}),
%
\begin{equation}
	\tilde g(t) = g_i (1-t/\tau) + i \delta_f t/\tau, \qquad 0 \le t \le \tau.
	\label{eq:gtilde_protocol1}
\end{equation}
%
The time-dependent Schr\"{o}dinger equation of the Hamiltonian $\mathcal{H}_k(t)$ in Eq. (\ref{eq:BdG_matrix}) 
is exactly solvable in the presence of complex field given by Eq.~(\ref{eq:gtilde_protocol1}) (see Appendix~\ref{app:weber}).
%
%
\begin{figure}[!t]
        \centering
       \includegraphics[scale=.48]{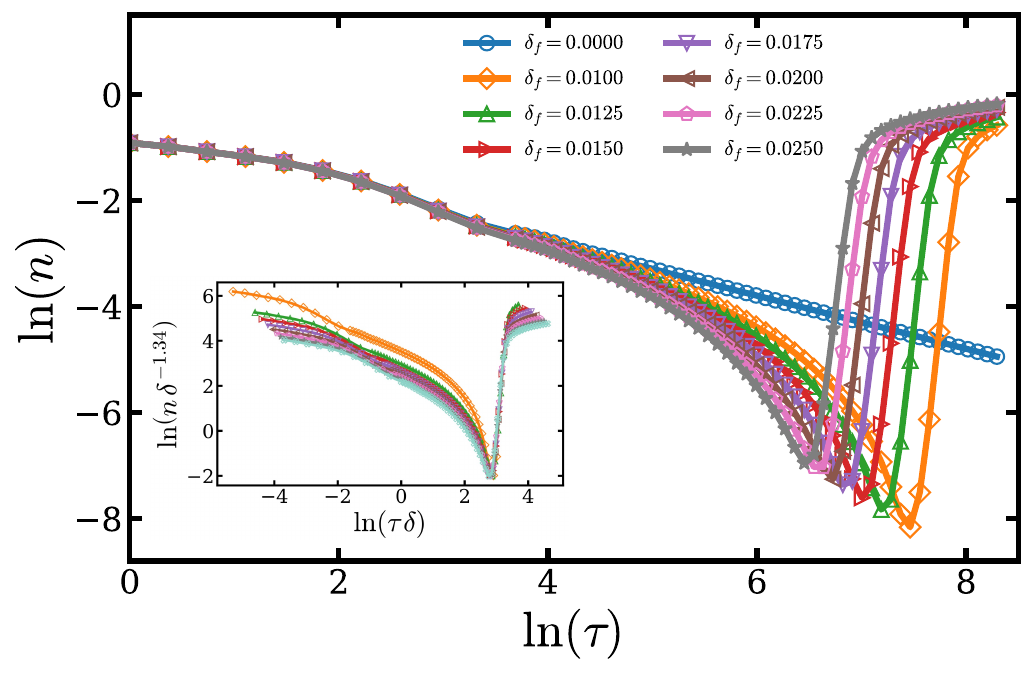}
        \caption{(Color online) The density of excitations vs annealing 
        time for quench path ($2$) is plotted in the main panel, where $J=0.5$, $g_i=4$, and $N=1024$ for various values of dissipation ($\delta_f$)
        (Inset) The scaling behavior of $n(\tau)$ near the minimum is observed by collapsing all curves of various $\delta$ by plotting $n(\tau)/\delta^{\gamma_2}$ vs $\tau/\delta^{-\alpha_2}$.}
\label{fig:P2_n_heatmap}
\end{figure}
%

The density of excitations as a result of Eqs.~\eqref{eq:Pk_num_correct} and \eqref{eq:n_def}, is plotted versus annealing time in Fig.~\ref{fig:P2_n_heatmap} for quench protocol in Eq.~(\ref{eq:gtilde_protocol1}). 
As observed, in the absence of dissipation ($\delta=0$) the defect density follows the KZ scaling $n \sim \tau^{-1/2}$ for $\ln(\tau)>3.5$~\cite{Dziarmaga2005} and
the effect of dissipation is invisible for $\ln(\tau)<3.5$.

We have identified an intriguing and counterintuitive phenomenon in the presence of dissipation beyond $\ln(\tau) > 3.5$.
Surprisingly, despite the ramp crossing the exceptional point, the defect density exhibits suppression faster than that predicted by the KZM for the interval $\ln(\tau) \in [3.5, \ln(\tau^{(2)}_{opt})]$ until reaching the minimum $n^{(2)}_{min}$ at $\tau^{(2)}_{opt}$. Subsequently, the defect density increases as $\tau$ increases, deviating from KZ scaling and indicating a crossover to an AKZ regime [$\ln(\tau) > \ln(\tau^{(2)}_{opt})$].

Our numerical results further demonstrate that the optimal annealing time at which the defect density is minimized is scaled with dissipation as $\tau^{(2)}_{opt} \sim \delta^{-\alpha_2}$ with $\alpha_2 = 1 \pm 0.05$. Furthermore, the data indicate that the minima of the defect densities scale with dissipation as $n^{(2)}_{min} \sim \delta^{\gamma_2}$ with $\gamma_2 = 1.34 \pm 0.05$.
The scaling behavior suggests that the defect density near its minimum is invariant for various dissipation parameters ($\delta$).  The inset of Figure~\ref{fig:P2_n_heatmap} illustrates this behavior, where we plot $n(\tau)/n^{(2)}_{min}$ and $\tau/\tau^{(2)}_{opt}$. The scaling function represents the promised universality of $n(\tau)$ in the face of dissipation.

%
\begin{figure}[!t]
\centerline{
\includegraphics[width=0.8\linewidth]{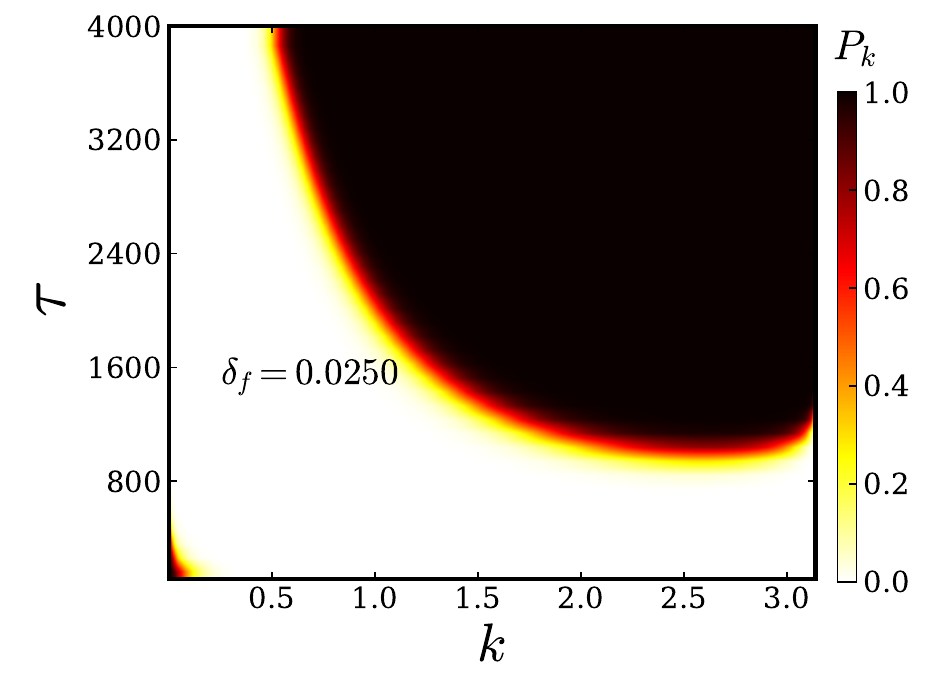}}
\caption{(Color online) The density plot of excitation probability at the end 
of quench $P_k(\tau)$ vs $k$ and $\tau$ for path~($2$) with the final dissipation value $\delta_f = 0.025$. The corresponding plot for unitary evolution $\delta=0$ is similar to left panel of Fig.~\ref{fig:P1_heatmap}.
}
\label{fig:P2_heatmap}
\end{figure}



To understand the origin of a suppression faster than the conventional KZ scaling and also AKZ behavior, the density plot of excitation probability has been depicted in Figure~\ref{fig:P2_heatmap} for $\delta=0.025$.
It is clear that, for a small annealing time [$\ln(\tau)<3.5$], the excitation probability is non-zero around the gap-closing mode. However, there is a range of annealing time $\ln(\tau) \in [3.5,\ln(\tau^{(2)}_{opt})]$ over which the excitation probability is small for all $k$ modes. While beyond this annealing time interval $\ln(\tau)>\ln(\tau^{(2)}_{opt})$, the excitation probability is significant for modes away from the gap-closing mode.
Therefore, the faster suppression of defect density with increasing $\tau$ than that predicted by the Kibble-Zurek mechanism can be attributed to the very small excitation probability over the annealing time interval $\ln(\tau) \in [3.5,\ln(\tau^{(2)}_{opt})]$. However, the AKZ behavior of the defect density originates from the fact that the range of modes over which the system undergoes nonadiabatic evolution enhances by increasing $\tau$ for $\ln(\tau)>\ln(\tau^{(2)}_{opt})$.

\subsection{The quench path ($3$): crossing two exceptional points}
\label{subsec:protocol3}    

In the quench path ($3$), the real part of the complex field is linearly ramped down from $g(0)=g_i>1$ at $t_i=0$ to $g(\tau)=g_f<-1$ at $t_f=\tau$ while the imaginary part remains a constant, i.e.,
%
\begin{equation}
	\tilde g(t) = (g_f - g_i)  t/\tau + g_i + i \delta, \qquad 0 \le t \le \tau.
	\label{eq:gtilde_protocol3}
\end{equation}
%
In this case, the system crosses two exceptional points during the quench.

\begin{figure}[!t]
\centerline{
\includegraphics[width=0.95\linewidth]{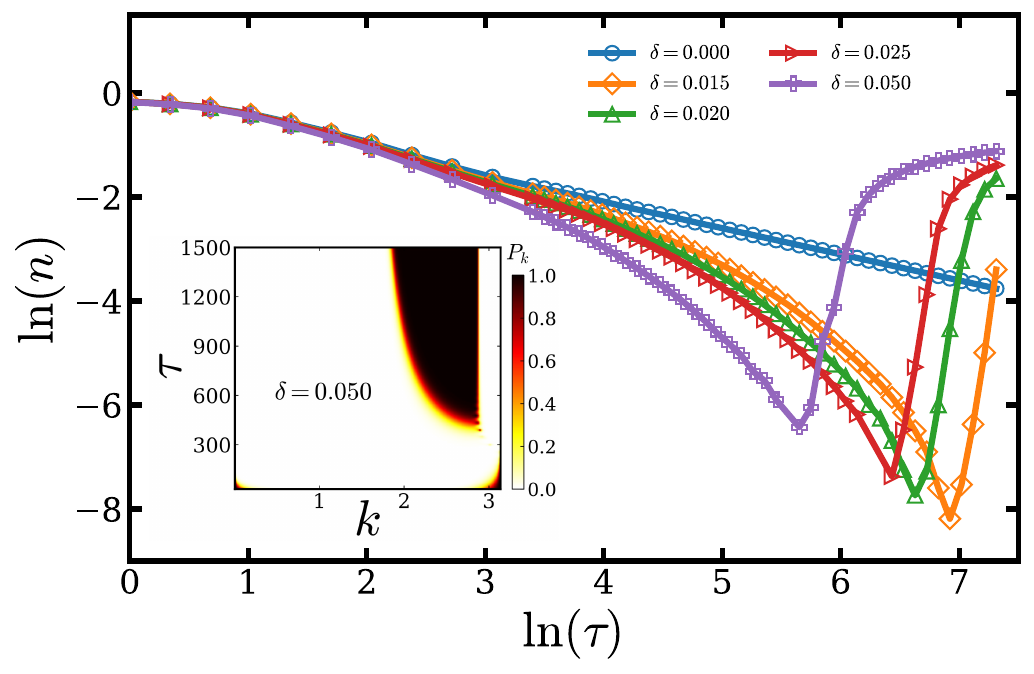}}
        \caption{(Color online) Density of excitation vs annealing time 
        for the quench path ($3$) with $J=0.5$, $g_i=2$, $g_f=-2$, $N=1024$ and several fixed dissipation ($\delta$). 
        A larger value of $\delta$ suppresses the excitations. Inset: the density
        plot of the probability of excited modes vs $k$ and $\tau$ and $\delta=0.05$ 
        demonstrates that for longer annealing times, modes other than those 
        corresponding to exceptional points ($k \sim 0$ and $k \sim \pi$) contribute significantly.}
        \label{fig:P4_n_heatmap}
    \end{figure}

The density of defects given by Eqs.~\eqref{eq:n_def} and \eqref{eq:Pk_num_correct}, has been plotted in Fig.~\ref{fig:P4_n_heatmap} versus the annealing time for different values of dissipation.
As observed, in the presence of fixed dissipation, the behavior of defect density for a quench of the transverse magnetic field across the exceptional point is remarkable. For a fast quench (small annealing time), the effects of dissipation on defect density are negligible. As the annealing time increases, the resulting defect density exhibits a markedly faster suppression than that predicted by the KZM until it reaches its minimum at the optimum annealing time $\tau^{(3)}_{opt}$.
Beyond the optimum annealing time $\tau>\tau^{(3)}_{opt}$, dissipation-induced excitations dominate the nonadiabatic dynamics, leading to the growth of defect density with the ramp time scale, which is indicative of a crossover to an AKZ regime.

Inset of Figure~\ref{fig:P4_n_heatmap} shows the density plot of excitation probability at the end of the quench. In the quench path ($3$), the modes corresponding to both exceptional points ($k\sim 0$ and $k\sim \pi$) are dominantly populated for short annealing times. However, there is a range of annealing time over which the excitation probability is negligible for all $k$ modes and beyond this interval, the excitation probability of modes away from the gap-closing modes is notable.
Analogous to the quench path ($2$), the deviation of defect density from the standard Kibble-Zurek scaling (suppression and enhancement) can be understood in terms of excitation probability. The unusual fast decay arises from the vanishing excitation probability over a range of annealing time for all $k$ modes even at the gap-closing modes. The AKZ behavior originates from supplementary excitation facilitated by dissipation over a broad range of allowed modes, specifically away from the gap-closing modes.   


In addition, the scaling behavior of the minimum of defect density and optimal annealing time versus dissipation is given as $n^{(3)}_{\text{min}} \sim \delta^{\gamma_3}$ and $\tau^{(3)}_{\text{opt}} \sim \delta^{-\alpha_3}$ with $\gamma_3 = 1 \pm 0.05$ and $\alpha_3 = 1 \pm 0.05$.

%
%

We have also studied other paths within linear protocols with non-Hermitian terms, namely path (4), which avoids crossing an exceptional point and also the reverse of path (1) (see Appendix~\ref{app:quech}). In the aforementioned cases, the presence of loss does not lower the excitation density, while the feature of exciting other modes than the gap-closing ones is present.


\section{Summary and Conclusions}\label{sec:conclusions}
We revisit non-Hermitian quantum annealing in the one-dimensional ferromagnetic Ising chain from a momentum-resolved perspective. We provide a systematic evaluation of intrinsic transition probabilities and the resulting defect density. Our approach retains the full Weber-function analytic solution for each momentum sector, fixes the mode-dependent coefficients from the initial conditions, and performs the momentum sum without any truncation. This reveals the contribution of broad momentum sectors to the final defect density in the non-Hermitian regime and delineates parameter ranges where gap-closing mode approximations are reliable. We then demonstrate how the corrected momentum-resolved treatment reshapes the predicted dependence of the kink density on dissipation strength and annealing time, and discuss implications for assessing the performance of non-Hermitian annealing protocols in integrable settings.

We have studied several protocols with a non-Hermitian Hamiltonian that mimics the dissipation in an open quantum system. Although the quench from a large magnetic field and small dissipation to a zero field and dissipation does not lower the defect density, it rather abruptly enhances such densities. We proposed alternative quench paths, such as path ($2$), to drastically diminish defect formations. Path ($2$) initiates from a large magnetic field and zero dissipation, gradually increasing dissipation until it ends with a finite dissipation and a zero field. Even a quench with constant dissipation would suppress the excitation density as shown in path ($3$). In all of these paths we considered with a non-Hermitian term, an anti-Kibble-Zurek behavior is observed. The density of excitations decreases by increasing annealing time before reaching a minimum, which defines the corresponding optimal annealing time. The optimal annealing time scales power law with the strength of non-Hermition parameter with a negative exponent ($\simeq -1$).

We have studied two additional annealing paths (presented in Appendix~\ref{app:quech}), a quench that avoids crossing an exceptional point and the reverse direction of path ($1$). In both cases, we observe that almost all modes contribute to excite defects, especially for long annealing time, which is in agreement with our finding for non-Hermitian quench dynamics.

Despite its appealing structure, translating exact mode-resolved dynamics into asymptotic predictions for defect densities and performance metrics in the thermodynamic limit is challenging. In Hermitian ramps across the Ising critical point, the gap-closing modes ($k^*$) dominate defect production because the minimal gap is controlled by $k \sim k^*$, allowing controlled $k$ expansions and Gaussian approximations to the excitation kernel. In contrast, for non-Hermitian ramps, the relevant spectral landscape can be displaced in momentum space, leading to complex gaps and exceptional-point physics that shift the locus of minimal adiabatic protection away from $k^*$. This can qualitatively reshape how different momentum sectors contribute to normalized observables. Consequently, approximations that restrict the dynamics or final observables to a window of gap-closing modes are not generally justified in non-Hermitian annealing, even when the underlying model remains mode-separable.

\begin{acknowledgments}
	This work was supported by the Iran National Science Foundation (INSF) under Grant No.~4037050. The authors would like to acknowledge financial support from Sharif University of Technology through Grant No.~G4040219.
\end{acknowledgments}

\appendix
\section{Jordan-Wigner transformation of the non-Hermitian Ising chain}
\label{app:JW}

In the appendix we derive in detail the transformed
form of the spin Hamiltonian Eq. (\ref{eq:H_spin}) using the Jordan-Wigner (JW) transformation ~\cite{Lieb1961,Pfeuty1970}, i.e., 
%
\begin{subequations}
	\label{eq:JW_map_app}
	\begin{align}
		\sigma_n^{x} &= 1 - 2 c_n^\dagger c_n,
		\label{eq:JW_sigma_x_app}
		\\
		\sigma_n^{y} &= i\,(c_n^\dagger - c_n)\,
		S_n,
		\label{eq:JW_sigma_y_app}
		\\
		\sigma_n^{z} &= - (c_n + c_n^\dagger)\,
		S_n,
		\label{eq:JW_sigma_z_app}
	\end{align}
\end{subequations}
%
where
%
\begin{equation}
	S_n \equiv \prod_{m<n} (1-2 n_m),
	\qquad
	n_m \equiv c_m^\dagger c_m ,
	\label{eq:Sn_def_app}
\end{equation}
%
and the fermions satisfy the canonical anticommutation relations
%
\begin{equation}
	\{c_m, c_n^\dagger\} = \delta_{mn},\qquad
	\{c_m, c_n\} = \{c_m^\dagger,c_n^\dagger\} = 0.
	\label{eq:CAR_app}
\end{equation}
%
It should be emphasized that, the string operator $S_n$ encodes the non-local Jordan-Wigner phase
and will be responsible for the subtle boundary term discussed below.


%
%

To this aim we split the Hamiltonian Eq. (\ref{eq:H_spin}) into three parts
%
\begin{equation}
	H = H_g + H_\delta + H_{zz}.
	\label{eq:H_split_app}
\end{equation}
%





\paragraph*{The transverse field term $H_g$.}
Contribution from the transverse-field is given as
%
\begin{equation}
	H_g = -\frac{J}{2} g \sum_{n=1}^{N} \sigma_n^{x}.
	\label{eq:Hg_def_app}
\end{equation}
%
Using Eq.~\eqref{eq:JW_sigma_x_app}, $H_g$ is transformed to  
%
\begin{equation}
	H_g
	= -\frac{J}{2} g \sum_{n=1}^{N} (1 - 2 c_n^\dagger c_n).
	\label{eq:Hg_JW_app}
\end{equation}
%
As clear, $H_g$ produces a constant shift \(-JgN/2\) and a local ``chemical potential'' term \(Jg\, n_n\) in the fermionic language.

\paragraph*{The dissipative term $H_\delta$.}
The dissipative contribution is given by  
%
\begin{equation}
	H_\delta = -\frac{J}{2} i\,2\delta \sum_{n=1}^{N} \sigma_n^{-}\sigma_n^{+},
	\label{eq:Hdelta_def_app}
\end{equation}
%
where
%
\begin{equation}
	\sigma_n^{-}\sigma_n^{+}
	= \frac{1}{4}
	(\sigma_n^{z} - i\sigma_n^{y})(\sigma_n^{z} + i\sigma_n^{y})
	= \frac{1}{2}\left(\mathbb{1} + \sigma_n^{x}\right).
	\label{eq:sigmasmsp_to_sigmax_app}
\end{equation}
%

Substituting \eqref{eq:JW_sigma_x_app} into \eqref{eq:sigmasmsp_to_sigmax_app} results

%
%
\begin{equation}
	H_\delta
	= -i J \delta \sum_{n=1}^{N} (1-c_n^\dagger c_n).
	\label{eq:Hdelta_JW_step_app}
\end{equation}
%
For simplicity, we introduce complex field $\tilde g \equiv g + i\delta$ by combining Eqs.~\eqref{eq:Hg_JW_app} and \eqref{eq:Hdelta_JW_step_app},
%
\begin{equation}
	H_g + H_\delta
	= -\frac{J}{2}\sum_{n=1}^{N}
	\left(\tilde{g} + i\delta - 2\tilde{g}\, c_n^\dagger c_n\right).
	\label{eq:HgHdelta_JW_app}
\end{equation}
%
The first term in Eq.~\eqref{eq:HgHdelta_JW_app},
proportional to $(\tilde g + i\delta)$, is a site-independent complex
energy shift, while the second contribution, proportional to
$-2\tilde g\,c_n^\dagger c_n$, plays the role of an effective on-site
potential in the fermionic language.

\paragraph*{Ising interaction $H_{zz}$.}

The Ising interaction is given as
%
\begin{equation}
	H_{zz} =  H_{zz}^{\text{blk}} + H_{zz}^{\text{bnd}}\\
\no
 H_{zz}^{\text{blk}}= -\frac{J}{2}\sum_{n=1}^{N-1} \sigma_n^{z}\sigma_{n+1}^{z},\\
 \no
H_{zz}^{\text{bnd}}= -\frac{J}{2}\sigma_N^{z}\sigma_{N+1}^{z}
\qquad \sigma_{N+1}^{z} \equiv \sigma_1^{z},
	\label{eq:Hzz_def_app}
\end{equation}
%
in which the Hamiltonian is split into two terms, the bulk Hamiltonian $H_{zz}^{\text{blk}}$ and the boundary term $H_{zz}^{\text{bnd}}$.

Using Eq.~\eqref{eq:JW_sigma_z_app}, i.e., 
%
\begin{equation}
	\sigma_n^{z}
	= - (c_n + c_n^\dagger) S_n,
	\qquad
	\sigma_{n+1}^{z}
	= - (c_{n+1} + c_{n+1}^\dagger) S_{n+1},
	\label{eq:sigmaz_bulk_app}
\end{equation}
%
and the following equation 
%
\begin{equation}
	S_{n+1}
	= S_n (1-2n_n),
	\label{eq:Snplus1_app}
\end{equation}
the ``bulk'' Ising term in the fermion language is given as
%
\begin{equation}
	H_{zz}^{\text{blk}}
	= -\frac{J}{2}
	\sum_{n=1}^{N-1}
	\left(
	c_n^\dagger c_{n+1}^\dagger
	+ c_{n+1} c_n
	+ c_n^\dagger c_{n+1}
	+ c_{n+1}^\dagger c_n
	\right).
	\label{eq:Hzz_bulk_JW_app}
\end{equation}
%

Moreover, the boundary term in fermion language is obtained as
%
\begin{equation}
	\sigma_N^{z}\sigma_1^{z}
	=
	(c_N + c_N^\dagger) S_N (c_1 + c_1^\dagger).
	\label{eq:sigmazsigmaz_boundary_step1_app}
\end{equation}
%
Introducing the fermion-parity operator~\cite{Lieb1961,Pfeuty1970}
%
\begin{equation}
	P \equiv \prod_{m=1}^{N} (1-2n_m),
	\label{eq:P_def_app}
\end{equation}
%
and using the equation 
%
\begin{equation}
	S_N = P(1-2n_N),
	\label{eq:SN_in_terms_of_P_app}
\end{equation}
%
%
%
%
The boundary contribution to the Ising term follows as
%
\begin{equation}
	H_{zz}^{\text{bnd}}
	= -\frac{J}{2}\sigma_N^{z}\sigma_1^{z}
	= -\frac{J}{2}(c_N^\dagger - c_N)P(c_1 + c_1^\dagger).
	\label{eq:Hzz_boundary_JW_app}
\end{equation}
%
In the even parity sector, i.e., $P=+1$, which forces to impose anti-periodic boundary condition ($c_{N+1}=-c_1$) in
the fermion chain and consequently leading to $k = 2\pi p/N$, with $p=1/2,3/2,\ldots, N-1/2$
the boundary term is given as 
%
\begin{equation}
	H_{zz}^{\text{bnd}}
	= -\frac{J}{2}(c_N^\dagger - c_N)(c_1 + c_1^\dagger).
	\label{eq:Hzz_boundary_pm_app}
\end{equation}
Finally, the full fermion Hamiltonian in the even parity sector is obtained as 
%
\begin{equation}
	\begin{aligned}
		H^{\pm}
		= -\frac{J}{2}\sum_{n=1}^{N}
		\biggl[
		&\,c_n^\dagger c_{n+1}
		+ c_{n+1}^\dagger c_n
		+ c_n^\dagger c_{n+1}^\dagger
		+ c_{n+1} c_n
		\\
		& - 2\tilde g\, c_n^\dagger c_n
		+ \tilde g + i\delta
		\biggr],
	\end{aligned}
	\label{eq:H_pm_final_app}
\end{equation}
%
with $\tilde g = g + i\delta$. 
It is worth mentioning that, in the thermodynamic limit, the even and odd parity sectors yield the identical results.

\section{Diagonalization of the non-Hermitian Hmiltonian}
\label{app:Bogoliubov}

We use the Bogoliubov transformation that diagonalizes the non-Hermitian Bogoliubov-de Gennes block appearing in the $(k,-k)$ sector~\cite{Nesterov2013}
%
\begin{equation}
	\label{B1}
	\bar{H}_k = \Psi_k^\dagger \mathcal{H}_k \Psi_k ,
		\qquad
	\Psi_k =
	\begin{pmatrix}
		c_k \\
		c_{-k}^{\dagger}
	\end{pmatrix},
\end{equation}
%
with
%
\begin{equation}
	\mathcal{H}_k =
	\begin{pmatrix}
		\Delta_k & \Omega_k \\
		\Omega_k & -\Delta_k
	\end{pmatrix},
	\label{BdGmatrix}
\end{equation}
%
where $\Delta_k = J(\tilde g-\cos k)$, and $\Omega_k = J\sin k$.
It should be mentioned that, since the constant term in the Hamiltonian is just the energy shift, we ignore the constant term in Nambu spinor form.

Since $\tilde g$ is complex, the Hamiltonian $\mathcal{H}_k$ is generally non-Hermitian and complex symmetric, $\mathcal{H}_k^{T}=\mathcal{H}_k$.\\

\paragraph*{Right and left Bogoliubov transformations.}

Since the Hamiltonian is non-Hermitian, we introduce the right quasiparticle operators~\cite{Brody2014,Bergholtz2021,Ashida2020,Deng2025} $\Phi_k=(\gamma_k,\gamma_{-k}^\dagger)^T$
defined by $\Psi_k = W_k \Phi_k$ with
%
\begin{equation}
	W_k =
	\begin{pmatrix}
		u_k & -v_k \\
    	v_k & u_k  
	\end{pmatrix}.
	\label{Wk}
\end{equation}
Moreover, the left quasiparticle operators
\begin{equation}
	\Phi_k^\dagger =
	(\tilde\gamma_k^\dagger,\tilde\gamma_{-k})
\end{equation}
%
are introduced via the transformation $\Psi_k^\dagger =	\Phi_k^\dagger \widetilde W_k^{T},$
with
%
\begin{equation}
	\widetilde W_k =
	\begin{pmatrix}
	\tilde u_k &	-\tilde v_k   \\
	\tilde v_k  &   \tilde u_k  
	\end{pmatrix}.
	\label{Wtilde}
\end{equation}
%
It should be mentioned that, we have applied the biorthogonality condition of the quasiparticle operators $\widetilde W_k^{T} W_k = \mathbb{1}$, 
which results in $\tilde u_k u_k + \tilde v_k v_k = 1,	\, \tilde u_k v_k = \tilde v_k u_k$.\\

Using the transformations above the Hamiltonian in Eq. (\ref{B1}) is given as
%
\begin{equation}
	\bar{H}_k
	=
	\Phi_k^\dagger
	\left(
	\widetilde W_k^{T}
	\mathcal{H}_k
	W_k
	\right)
	\Phi_k .
\end{equation}
%
By defining 
%
\begin{equation}
	\widetilde W_k^{T}\mathcal{H}_k W_k =
	\begin{pmatrix}
		(\tilde H_k)_{11} & (\tilde H_k)_{12} \\
		(\tilde H_k)_{21} & (\tilde H_k)_{22}
	\end{pmatrix},
\end{equation}
%
and putting the off-diagonal terms 
%
\begin{equation}
	(\tilde H_k)_{12}
	=
	-\Omega_k(\tilde u_k u_k-\tilde v_k v_k)
	+
	2\Delta_k \tilde u_k v_k, 
\end{equation}
%
$(\tilde H_k)_{12}=0 $ equals zero, the diagonalization condition is obtained as 
%
\begin{equation}
	\Omega_k(\tilde u_k u_k-\tilde v_k v_k)
		=
		2\Delta_k \tilde u_k v_k.
	\label{condition_general}
\end{equation}
%
This condition can be also obtained using the Bogoliubov transformation.

\paragraph*{Right and left eigenvectors.}

The right and left eigenvectors of the non-Hermitian BdG matrix in Eq. (\ref{BdGmatrix}) are obtained as~\cite{Brody2014,Bergholtz2021,Ashida2020,Deng2025} 
%
\begin{equation}
	\mathcal{H}_k |u_\lambda(k)\rangle
	=
	E_\lambda(k) |u_\lambda(k)\rangle ,
\end{equation}
%
%
\begin{equation}
	\langle \tilde u_\lambda(k)|
	\mathcal{H}_k
	=
	E_\lambda(k)
	\langle \tilde u_\lambda(k)|,
\end{equation}
%
where $E_\pm(k)=\pm E_k ,\quad E_k=\sqrt{\Delta_k^2+\Omega_k^2}$ is the eigenvalues of the non-Hermitian Hamiltonian in Eq. (\ref{BdGmatrix}). 
The right $|u_\lambda(k)\rangle$ and the left $\langle \tilde u_\lambda(k)|$ eigenvectors should be satisfy the biorthonormality condition
%
\begin{equation}
	\langle \tilde u_\lambda(k)|u_{\lambda'}(k)\rangle
	=
	\delta_{\lambda\lambda'}.
\end{equation} 
%

By defining the ground and excited states of the right and left eigenvectors as
%
\begin{align}
	|u_+(k)\rangle
	&=
	\begin{pmatrix}
		u_k \\
		v_k
	\end{pmatrix},
	\label{app:rightvec+}
	\\
	|u_-(k)\rangle
	&=
	\begin{pmatrix}
		- v_k \\
		u_k
	\end{pmatrix},
	\label{app:rightvec-}
\end{align}
%
%
\begin{align}
	\langle \tilde u_+(k)|
	&=
	\begin{pmatrix}
		\tilde u_k & \tilde v_k
	\end{pmatrix},
	\\
	\langle \tilde u_-(k)|
	&=
	\begin{pmatrix}
		-\tilde v_k & \tilde u_k
	\end{pmatrix}.
	\label{app:leftvec}
\end{align}
and applying the complex-symmetric condition $\mathcal{H}_k^{T}=\mathcal{H}_k$ (which impose 
$\tilde u_k = u_k,\,\tilde v_k = v_k$), Eq. (\ref{condition_general}) is obtained as 
%
\begin{equation}
	\Omega_k(u_k^2-v_k^2)=2\Delta_k u_k v_k .
\end{equation}
%
Defining $u_k=\cos(\theta_k/2),\,v_k=\sin(\theta_k/2)$ as a normalization condition
one finds
%
\begin{equation}
	\tan(\theta_k)=\frac{\Omega_k}{\Delta_k},
	\label{eq:theta-tan}
\end{equation}
%
which results
%
\begin{align}
	\cos(\theta_k) &= \frac{\tilde{g} - \cos (k)}{\sqrt{\tilde{g}^2 - 2\tilde{g} \cos (k) + 1}}, \label{eq:theta-cos} \\
	\sin(\theta_k) &= \frac{\sin (k)}{\sqrt{\tilde{g}^2 - 2\tilde{g} \cos (k) + 1}} \label{eq:theta-sin}.
\end{align}
%
\section{Time dependent Schr\"{o}dinger equation}
\label{app:weber}

Equation~\eqref{eq:Sch_2level} is the time dependent Schr\"{o}dinger equation of the driven two-level system with constant
off-diagonal coupling $\Omega_k$ and a generally complex, time-dependent
detuning $\Delta_k(t)$.
Using the coupled first-order differential equations in Eq.~\eqref{eq:Sch_2level},
%
\begin{align}
	i\dot C_1 &= \Delta_k(t)\,C_1 + \Omega_k\,C_2, \label{eq:A1}\\
	i\dot C_2 &= \Omega_k\,C_1 - \Delta_k(t)\,C_2, \label{eq:A2}
\end{align}
%
$C_2(t)$ can be calculated from Eq.~\eqref{eq:A1},i.e.,
%
\begin{equation}
	C_2 = \frac{i\dot C_1-\Delta_k C_1}{\Omega_k}.
	\label{eq:C2_from_C1}
\end{equation}
%
Then, substituting Eq. (\ref{eq:C2_from_C1}) in differentiated of Eq. (\ref{eq:A1}), we get the second-order differential equation for $C_1$
%
\begin{equation}
	\ddot C_1 + \Big(\Omega_k^2+\Delta_k^2+i\dot\Delta_k\Big)C_1=0.
	\label{eq:C1_second_order_exact}
\end{equation}
%
The exact solution of second order differential equation is available by linearizing  $\Delta_k(t)$ around the
(complex) avoided-crossing time $t_0$ as $\Delta_k(t_0)=0$~\cite{VitanovGarraway1996,Vitanov1999,Zener1932,Pechukas1976}. 
Assuming analyticity of $\tilde g(t)$ in a neighborhood of $t_0$, the detuning
admits the Taylor expansion
%
\begin{align}
	\Delta_k(t)
	&= \Delta_k(t_0)
	+ \dot\Delta_k(t_0)(t-t_0)
	+ \cdots
	\nonumber\\
	&= \dot\Delta_k(t_0)(t-t_0)
	+ \mathcal{O}\!\left((t-t_0)^2\right),
	\label{eq:Delta_Taylor}
\end{align}
%
where we used $\Delta_k(t_0)=0$, and keeping only the linear term yields the local Landau-Zener problem~\cite{Vitanov1999,VitanovGarraway1996}
%
\begin{equation}
	\Delta_k(t)\simeq a_k (t-t_0),
	\qquad
	a_k \equiv \dot\Delta_k(t_0)=J\dot{\tilde g}(t_0).
	\label{eq:Delta_linear_app}
\end{equation}
%
Within this approximation $\dot\Delta_k = a_k$ is constant, and
Eq.~\eqref{eq:C1_second_order_exact} becomes
%
\begin{equation}
	\ddot C_1
	+ \Big[\Omega_k^2 + a_k^2 (t-t_0)^2 + i a_k \Big] C_1 = 0.
	\label{eq:C1_second_order_linearized}
\end{equation}
%
Introduce the dimensionless variable
%
\begin{equation}
	T=\sqrt{a_k}\,(t-t_0),
	\label{eq:T_map_app}
\end{equation}
%
where $\sqrt{a_k}$ denotes a fixed choice of square-root branch. Dividing
Eq.~\eqref{eq:C1_second_order_linearized} by $a_k$ yields
%
\begin{equation}
	\frac{d^2 C_1}{dT^2}+\left[T^2+i+\frac{\Omega_k^2}{a_k}\right]C_1=0.
	\label{eq:C1_T_equation}
\end{equation}
%
To bring this into the canonical parabolic-cylinder form, define
%
\begin{equation}
	z \equiv \sqrt{2}\,e^{-i\pi/4}T,
	\qquad
	\omega_k\equiv \frac{\Omega_k}{\sqrt{a_k}},
	\qquad
	\nu_k\equiv \frac{i}{2}\omega_k^2.
	\label{eq:z_omega_nu_app}
\end{equation}
%
A straightforward change of variables shows that Eq.~\eqref{eq:C1_T_equation}
is equivalent to
%
\begin{equation}
	\frac{d^2 C_1}{dz^2}+\left(\nu_k+\frac{1}{2}-\frac{z^2}{4}\right)C_1=0,
	\label{eq:Weber_standard}
\end{equation}
%
whose independent solutions are the parabolic-cylinder functions $D_{\nu_k}(z)$
and $D_{\nu_k}(-z)$. Therefore \cite{Vitanov1999},
%
\begin{equation}
	C_1(z)=A_k\,D_{\nu_k}(z)+B_k\,D_{\nu_k}(-z).
	\label{eq:C1_general_app}
\end{equation}
%
\paragraph*{Expression for \texorpdfstring{$C_2$}{C2} in terms of Weber functions.}
\label{app:C2_from_weber}

Using Eq.~\eqref{eq:C2_from_C1} and
the chain rule, one can express $C_2$ in terms of $z$-derivatives of $C_1$.
The key identity for parabolic-cylinder functions is
%
\begin{equation}
	\frac{d}{dz}D_{\nu}(z)=\frac{z}{2}D_{\nu}(z)-D_{\nu+1}(z),
	\label{eq:D_derivative_identity}
\end{equation}
%
together with the recurrence
%
\begin{equation}
	D_{\nu+1}(z)=zD_{\nu}(z)-\nu D_{\nu-1}(z).
	\label{eq:D_recurrence}
\end{equation}
%
After a short algebra, one arrives at the compact form used in the main text \cite{Vitanov1999},
%
\begin{equation}
	C_2(z)=\frac{\omega_k\,e^{-i\pi/4}}{\sqrt{2}}
	\Big[-A_k\,D_{\nu_k-1}(z)+B_k\,D_{\nu_k-1}(-z)\Big].
	\label{eq:C2_general_app}
\end{equation}
%
Equations \eqref{eq:C1_general_app} and \eqref{eq:C2_general_app} provide the
Weber-function representation of the two-component state.

\paragraph*{Matching to general initial conditions.}
\label{app:weber_matching}

Let $t_i$ denote the initial time (in our simulations, $t_i=0$) and define
%
\begin{equation}
	T_i=\sqrt{a_k}(t_i-t_0),
	\qquad
	z_i=\sqrt{2}\,e^{-i\pi/4}T_i,
	\label{eq:zi_def}
\end{equation}
%
and similarly at the final time $t_f$ (in our simulations, $t_f=\tau$),
%
\begin{equation}
	T_f=\sqrt{a_k}(t_f-t_0),
	\qquad
	z_f=\sqrt{2}\,e^{-i\pi/4}T_f.
	\label{eq:zf_def}
\end{equation}
%
Evaluating Eqs.~\eqref{eq:C1_general_app}--\eqref{eq:C2_general_app} at $t_i$ gives
a linear system for the unknown coefficients $(A_k,B_k)$:
%
\begin{equation}
	\begin{pmatrix}
		D_{\nu_k}(z_i) & D_{\nu_k}(-z_i)\\[1mm]
		-\eta_k D_{\nu_k-1}(z_i) & \eta_k D_{\nu_k-1}(-z_i)
	\end{pmatrix}
	\begin{pmatrix}
		A_k\\ B_k
	\end{pmatrix}
	=
	\begin{pmatrix}
		C_1(t_i)\\ C_2(t_i)
	\end{pmatrix},
	\label{eq:matching_linear_system}
\end{equation}
%
where $\eta_k \equiv \frac{\omega_k\,e^{-i\pi/4}}{\sqrt{2}}$.\\

Here $C_1(t_i)$ and $C_2(t_i)$ denote the components of the instantaneous
ground-state eigenvector of $\mathcal{H}_k(t_i)$ in the even-parity basis
$\{|0\rangle, c_k^\dagger c_{-k}^\dagger|0\rangle\}$.
In our numerical implementation these are obtained by direct diagonalization
of Eq.~\eqref{eq:BdG_matrix} at $t_i$.

Solving Eq.~\eqref{eq:matching_linear_system} yields $(A_k,B_k)$ uniquely
provided the determinant is nonzero. In practice, we compute the determinant
explicitly and solve via Cramer's rule to preserve numerical stability in
high precision.

Once $(A_k,B_k)$ are known, the final amplitudes are obtained by substituting
$z=z_f$ into Eqs.~\eqref{eq:C1_general_app} and \eqref{eq:C2_general_app}.


\section{Numerical solution: direct integration of the mode-resolved Schr\"odinger equation}
\label{subsec:numerics}

In addition to the Weber-function construction, we perform an independent numerical integration of the mode-resolved Schr\"odinger equation in order (i) to benchmark the analytical results and (ii) to treat general quench contours $\tilde g(t)$ in the complex plane. Throughout we set $\hbar=1$.

For each momentum mode $k$, the two-component state
\(
\psi_k(t)=(v_k(t),q_k(t))^{\mathsf T}
\)
evolves according to
\begin{equation}
	\frac{d}{dt}\psi_k(t)=-i\mathcal H_k(t)\psi_k(t),
	\label{eq:tdse_numeric}
\end{equation}
where $\mathcal H_k(t)$ is the non-Hermitian $2\times2$ BdG block defined in Eq.~\eqref{eq:BdG_matrix}. The equation is solved using an implicit stiff integrator with adaptive tolerance, and the Jacobian $-i\mathcal H_k(t)$ is supplied explicitly to ensure stability for long annealing times and in parameter regimes with strong non-unitary amplification or attenuation \cite{Ashida2020,Nesterov2013}.

Because the evolution is non-unitary for $\delta\neq0$, the physically meaningful mode occupation at the final time $t=\tau$ must be extracted using a properly normalized biorthogonal projection. At $t=\tau$, we diagonalize $\mathcal H_k(\tau)$,
\begin{equation}
	\mathcal H_k(\tau)\,|u_{n,k}\rangle
	=
	\lambda_{n,k}\,|u_{n,k}\rangle ,
\end{equation}
and construct the corresponding left eigenvectors $|\tilde u_{n,k}\rangle$ via the biorthonormal condition
$
	\langle \tilde u_{m,k}|u_{n,k}\rangle=\delta_{mn}.
$

The final state is expanded in the normalized right eigenbasis \cite{Brody2014,Deng2025},
\begin{equation}
	|\psi_k(\tau)\rangle
	=
	\sum_n c_{n,k}\,
	\frac{|u_{n,k}\rangle}
	{\sqrt{\langle u_{n,k}|u_{n,k}\rangle}},
\end{equation}
with coefficients
$
	c_{n,k}
	=
	\langle \tilde u_{n,k}|\psi_k(\tau)\rangle
	\sqrt{\langle u_{n,k}|u_{n,k}\rangle}.$

The intrinsic occupation weight of mode $n$ is then defined in a gauge-invariant manner as (see Appendix~\ref{app:biorth_prob}),
\begin{equation}
	O_{n,k}
	=
	\frac{|c_{n,k}|^2}
	{\sqrt{
			\langle u_{n,k}|u_{n,k}\rangle
			\langle \tilde u_{n,k}|\tilde u_{n,k}\rangle
	}},
	\label{eq:O_nk}
\end{equation}
and the corresponding mode occupation probability is
\begin{equation}
	P^{(\mathrm{num})}_{n,k}
	=
	\frac{O_{n,k}}
	{\sum_m O_{m,k}}.
	\label{eq:Pk_num_correct}
\end{equation}

For the excited mode (identified by the larger $\mathrm{Re}\,\lambda_{n,k}$), we denote
\(
P^{(\mathrm{num})}_{k}
=
P^{(\mathrm{num})}_{+,k}.
\)
In the Hermitian limit $\delta\to0$, this expression reduces to the standard adiabatic-basis occupation $|\langle u_{+,k}|\psi_k(\tau)\rangle|^2$. The numerical construction of the biorthonormal basis and the proof of gauge invariance of Eq.~\eqref{eq:Pk_num_correct} are summarized in Appendix~\ref{app:biorth_prob}~\cite{Brody2014,Deng2025}.

\section{Gauge-invariant biorthogonal projection}
\label{app:biorth_prob}

In this Appendix we summarize the numerical construction of the biorthonormal eigenbasis used in the simulations and provide a compact proof that the occupation probability defined in Eq.~\eqref{eq:Pk_num_correct} is invariant under the eigenvector rescaling (gauge) freedom inherent to non-Hermitian eigenproblems.

\paragraph*{Right/left eigenvectors and biorthonormalization.}
At the final time $t=\tau$, for each momentum mode $k$ we diagonalize the non-Hermitian BdG block $\mathcal H_k(\tau)$ and obtain the right eigenvectors $\{|u_{n,k}\rangle\}$ and eigenvalues $\{\lambda_{n,k}\}$,
\begin{equation}
	\mathcal H_k(\tau)\,|u_{n,k}\rangle=\lambda_{n,k}\,|u_{n,k}\rangle.
\end{equation}
Assuming $\mathcal H_k(\tau)$ is diagonalizable, we assemble the right eigenvectors as columns of a matrix
$V_R=[|u_{1,k}\rangle,\dots,|u_{N,k}\rangle]$.
The corresponding left eigenvectors $\{|\tilde u_{n,k}\rangle\}$ are chosen to satisfy the biorthonormality condition
\begin{equation}
	\langle \tilde u_{m,k}|u_{n,k}\rangle=\delta_{mn},
	\label{eq:biorth_cond_app}
\end{equation}
which we enforce numerically by constructing
\begin{equation}
	V_L=(V_R^{-1})^\dagger,
	\label{eq:VL_from_VR}
\end{equation}
so that $V_L^\dagger V_R=\mathbb{1}$ (up to machine precision). The columns of $V_L$ are the left eigenvectors $|\tilde u_{n,k}\rangle$ appearing in Eq.~\eqref{eq:biorth_cond_app}.
This construction avoids eigenvalue-matching ambiguities that can arise when separately diagonalizing $\mathcal H_k^\dagger(\tau)$.

Following Sec.~\ref{subsec:numerics}, we expand the final state in the normalized right and left eigenbases \cite{Deng2025},
\begin{eqnarray}
	|\psi_k(\tau)\rangle=\sum_n c_{n,k}\,
	\frac{|u_{n,k}\rangle}{\sqrt{\langle u_{n,k}|u_{n,k}\rangle}},
	\nonumber \\
	|\tilde\psi_k(\tau)\rangle=\sum_n c_{n,k}\,
	\frac{|\tilde u_{n,k}\rangle}{\sqrt{\langle \tilde u_{n,k}|\tilde u_{n,k}\rangle}},
	\label{eq:cn_app}
\end{eqnarray}
where $
	c_{n,k}=\langle \tilde u_{n,k}|\psi_k(\tau)\rangle
	\sqrt{\langle u_{n,k}|u_{n,k}\rangle}
$,
and define the intrinsic weight
\begin{equation}
	p_{n,k}(\tau)
	=
	\frac{\langle \tilde\psi_k(\tau)|u_{n,k}\rangle\,
		\langle \tilde u_{n,k}|\psi_k(\tau)\rangle}
	{\langle \tilde\psi_k(\tau)|\psi_k(\tau)\rangle\,
		\langle \tilde u_{n,k}|u_{n,k}\rangle}.
	\label{eq:pn_app_full}
\end{equation}
Numerically, we compute Eq.~\eqref{eq:Pk_num_correct}.

\paragraph*{Gauge freedom and invariance of the occupation probability.}
Because Eq.~\eqref{eq:biorth_cond_app} remains valid under the rescaling
$
|u_{n,k}\rangle \rightarrow |u'_{n,k}\rangle=\frac{1}{a_{n,k}}\,|u_{n,k}\rangle,
\,
|\tilde u_{n,k}\rangle \rightarrow |\tilde u'_{n,k}\rangle=a_{n,k}^*\,|\tilde u_{n,k}\rangle,
\label{eq:gauge_app}
$
with arbitrary nonzero complex $a_{n,k}$, a physically meaningful occupation must be invariant under this rescaling.

\paragraph*{Proof of gauge invariance.}
Under the rescaling, the coefficient in \eqref{eq:cn_app} transforms as
\begin{align}
	c'_{n,k}
	&=
	\langle \tilde u'_{n,k}|\psi_k(\tau)\rangle
	\sqrt{\langle u'_{n,k}|u'_{n,k}\rangle}\\
	&=
	\left(a_{n,k}\langle \tilde u_{n,k}|\psi_k(\tau)\rangle\right)
	\sqrt{\frac{1}{|a_{n,k}|^2}\langle u_{n,k}|u_{n,k}\rangle}
	\nonumber\\
	&=
	\frac{a_{n,k}}{|a_{n,k}|}\,
	\langle \tilde u_{n,k}|\psi_k(\tau)\rangle
	\sqrt{\langle u_{n,k}|u_{n,k}\rangle}
	=
	\frac{a_{n,k}}{|a_{n,k}|}\,c_{n,k}.
	\label{eq:cn_phase_app}
\end{align}
Hence $|c'_{n,k}|^2=|c_{n,k}|^2$. Moreover,
\begin{equation}
	\langle u'_{n,k}|u'_{n,k}\rangle =\frac{1}{|a_{n,k}|^2}\langle u_{n,k}|u_{n,k}\rangle, \nonumber
	\\
	\langle \tilde u'_{n,k}|\tilde u'_{n,k}\rangle =|a_{n,k}|^2\langle \tilde u_{n,k}|\tilde u_{n,k}\rangle,
\end{equation}
so the denominator in Eq.~\eqref{eq:O_nk} is invariant:
\begin{equation}
	\sqrt{\langle u'_{n,k}|u'_{n,k}\rangle\langle \tilde u'_{n,k}|\tilde u'_{n,k}\rangle}
	=
	\sqrt{\langle u_{n,k}|u_{n,k}\rangle\langle \tilde u_{n,k}|\tilde u_{n,k}\rangle}.
\end{equation}
Combining these results yields $O'_{n,k}=O_{n,k}$, and therefore
$P^{(\mathrm{num})}_{n,k}$ in Eqs.~\eqref{eq:Pk_num_correct} is gauge invariant under rescaling.
\hfill$\square$

\section{Quench Paths from the Origin to and Beyond the Exceptional Ring}
\label{app:quech}

\paragraph{Quench path ($4$): Avoiding exceptional point crossing}

\begin{figure}[!t]
\centerline{
\includegraphics[width=0.9\linewidth]{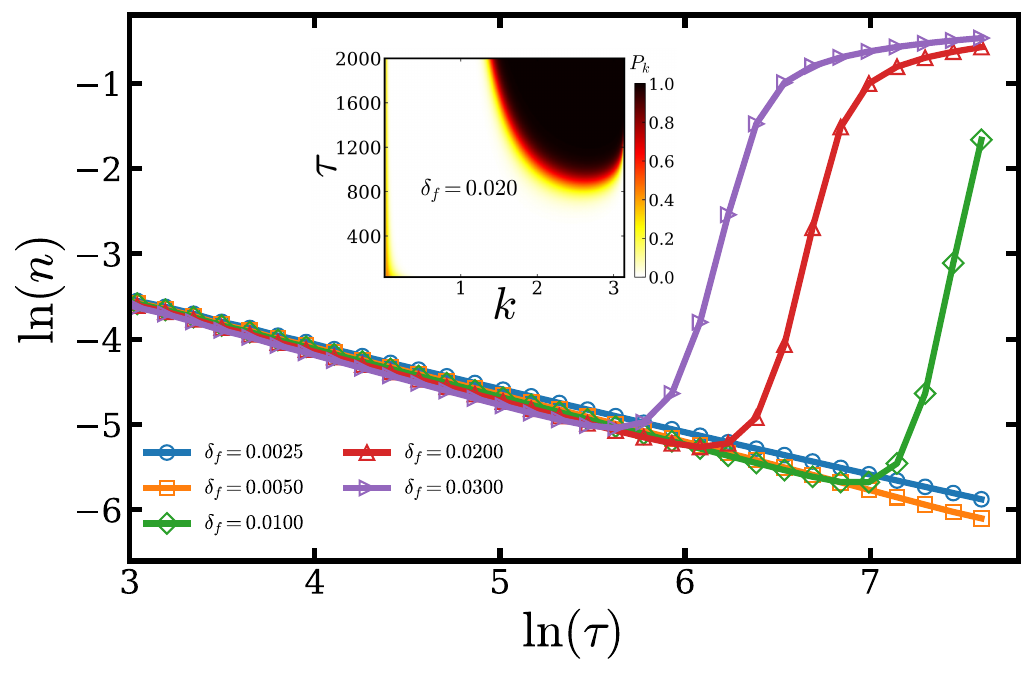}}
\caption{(Color online) The density of excitations vs the annealing time for quench path (4), with parameters $g_i = 0$, $\delta_i = 0$, and final dissipation strengths $\delta_f$ indicated in the legend. The final field is given by $g_f = \sqrt{1 - \delta_f^{\,2}}$, and the system size is $N = 1024$ with $J = 0.5$. (Inset) Excitation probability $P_k$ as a function of momentum $k$ and annealing time $\tau$ for $\delta_f = 0.020$, using the same parameters.}
\label{fig:nw3}
\end{figure}
In the quench path (4), the complex field is linearly swept from the origin $g_i=0, \delta_i=0$ 
to the complex value $g_f + i \delta_f$ on the exceptional ring, with quench protocol as 
%
\begin{equation}
	\tilde g(t)  = (g_f + i \delta_f) t/\tau, \qquad 0 \le t \le \tau,
	\label{eq:gtilde_protocol4}
\end{equation}
%
where $g_f > 0$ and $\delta_f > 0$ are real parameters chosen such that $g_{f}^{2} + \delta_{f}^{2} = 1$.
This quench path is marked as (4) in Figure~\ref{fig:phase_space}.

%
\begin{figure}[!t]
\centerline{
\includegraphics[width=0.9\linewidth]{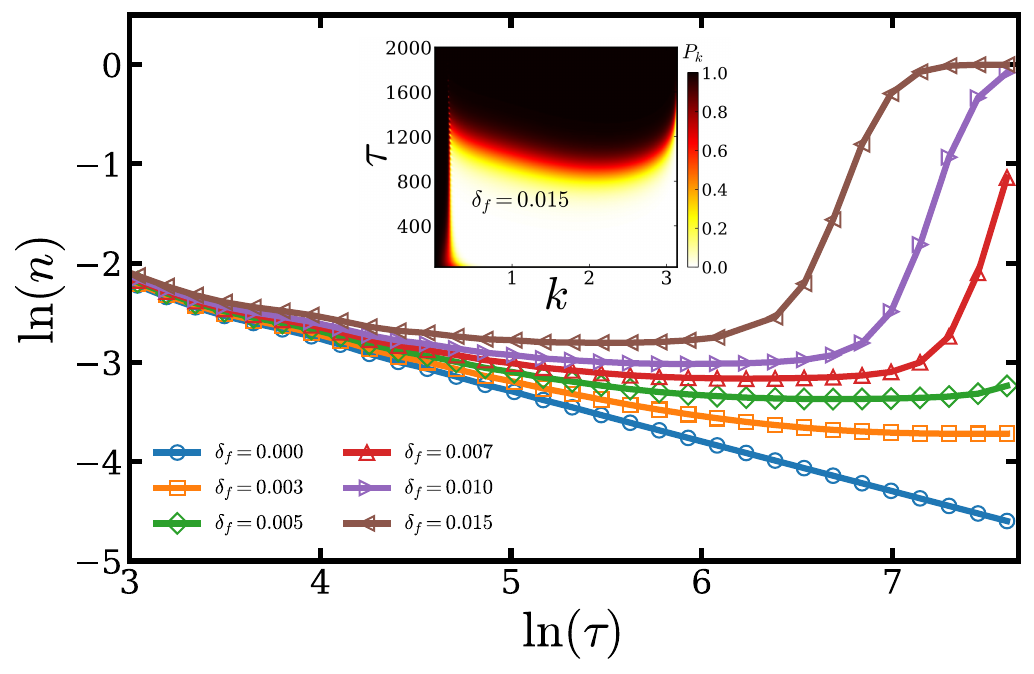}}
\caption{(Color online) Density of excitations vs annealing time for the reverse quench path (1) with parameters $g_i=0$, $\delta_i=0$, $g_f=4$, and the dissipation strengths $\delta_f$ shown in the legend. The system size is $N=1024$ with $J=0.5$. (Inset) Excitation probability $P_k$ as a function of $k$ and $\tau$ for $\delta_f = 0.015$, using the same parameters.}

\label{fig:nr1}
\end{figure}
%
In Fig.~\ref{fig:nw3}, the density of defects $n(\tau)$ for the quench protocol given in Eq. (\ref{eq:gtilde_protocol4}), has been plotted versus annealing time 
$\tau$ for different values of dissipation strength. It can be clearly seen that, for short annealing time, the effect of dissipation is invisible and 
the density of defects scales as a power law in agreement with the KZM prediction $n(\tau)\propto\tau^{-1/2}$ ($\delta=0$)~\cite{Dziarmaga2005}.
However, for longer annealing time, dissipation-induced effects dominate the nonadiabatic dynamics, leading to the growth of $n(\tau)$ 
with the annealing time $\tau$ which is indicative of a crossover to an AKZ regime.

The inset of Fig.~\ref{fig:nw3} plots the density plot of excitation probability $P_k(\tau)$ 
for $\delta_f=0.02$ at the end of a quench versus momentum and annealing time. 
The excitation probability is notable for a mode away from the gap-closing mode. As dissipation $\delta_f$ increases, 
the range of modes over which the transition probability is significant extended,  
while for $\delta_f=0$ the modes around the gap-closing mode are non-zero. 
This confirms the novel feature discussed in the main text regarding non-Hermitian dynamics.



\paragraph{Quench in reverse direction of quench path ($1$)}

In this path, the complex field is swept linearly from zero to complex value $g_f + i \delta_f$ outside the exceptional ring  ($\tau$) given by
%
\begin{equation}
	\tilde g(t)  = (g_f + i \delta_f) t/\tau, \qquad 0 \le t \le \tau,
	\label{eq:gtilde_protocol2}
\end{equation}
%
where $g_f>1$ and $\delta_f > 0$ are real parameters. 
In this annealing procedure, both magnetic field and dissipation are added to the system during the quench to reach the final 
large field and finite dissipation. 
The density of defects has been plotted versus $\tau$ in Fig.~~\ref{fig:nr1} for different values of dissipation. As seen for small annealing time the effects of dissipation 
is negligible while for large annealing time the defect densities increase by increasing the annealing time representing an anti-Kibble-Zurek behavior.

The density plot of excitation probability is depicted in the inset of Fig.~\ref{fig:nr1} for $\delta_f= 0.015$. 
The effects of dissipation reveals thought the notable excitation probability for broad range of allowed modes away from the gap-closing mode. 



%


\end{document}